%% file: main.tex
\documentclass[acmsmall,screen,nonacm]{acmart}

\input{macros}
\AtBeginDocument{%
  \providecommand\BibTeX{{%
    Bib\TeX}}}

\setcopyright{none}
\renewcommand\footnotetextcopyrightpermission[1]{}
\begin{document}


\title{\Sys: Synthesizing and Validating Access Control Policies}

\author{Ayesha Binte Mostofa}
\email{amostofa@umass.edu}
\affiliation{%
  \institution{University of Massachusetts Amherst}
  \city{Amherst}
  \state{Massachusetts}
  \country{USA}
}

\author{Gehao Zhang}
\email{gehaozhang@umass.edu}
\affiliation{%
  \institution{University of Massachusetts Amherst}
  \city{Amherst}
  \state{Massachusetts}
  \country{USA}
}

\author{Sam Stern}
\email{jstern@umass.edu}
\affiliation{%
  \institution{University of Massachusetts Amherst}
  \city{Amherst}
  \state{Massachusetts}
  \country{USA}
}

\author{Juan Zhai}
\email{juanzhai@umass.edu}
\affiliation{%
  \institution{University of Massachusetts Amherst}
  \city{Amherst}
  \state{Massachusetts}
  \country{USA}
}

\author{Pubali Datta}
\email{pdatta@umass.edu}
\affiliation{%
  \institution{University of Massachusetts Amherst}
  \city{Amherst}
  \state{Massachusetts}
  \country{USA}
}

\renewcommand{\shortauthors}{Mostofa et al.}

\ccsdesc[500]{Security and privacy~Access control}
\ccsdesc[500]{Software and its engineering~Software verification and validation}

\keywords{Access control, authorization policies, policy synthesis and repair, large language models, test oracle generation}




\begin{abstract}
Writing correct access control policies for modern software applications is difficult because it requires expertise in the authorization language and a thorough understanding of the application's authorization requirements. Existing synthesis and mining approaches assume explicit authorization inputs and a trusted reference policy. However, authorization logic is often distributed across code, comments, and documentation rather than captured in a single artifact, and developer-defined policies in real-world application repositories are often incomplete, outdated, or incorrect. We present \Sys, an LLM-aided pipeline for access control policy synthesis, repair, and validation. \Sys synthesizes policies through collecting repository evidence for authorization intent, using that evidence to guide an LLM to generate an initial policy. \Sys then uses official policy-language validation tools to identify and automatically repair errors. For semantic validation, a different LLM derives a behavioral oracle from the same repository evidence. The oracle consists of concrete authorization requests with expected allow or deny decisions, which \Sys executes against the generated policy and compares with the expected decisions. We manually assess the oracle on a statistically sized sample of 390 test cases. We evaluate \Sys on 228 real-world GitHub repositories that use Cedar,
OpenFGA, or Rego. In these repositories, 40.6--55.7\% of the repository
policies fail their language's validation tool. In contrast, \Sys produces 99.0--100\% valid policies after
repair. Their agreement with the oracle is 98.2--99.6\% for OpenFGA,
82.0--86.5\% for Cedar, and 76.3--80.2\% for Rego. Manual review confirms
91.5--96.9\% of the sampled test cases as correct. Synthesizing and
validating the policy for one repository takes about five minutes and costs
\$0.55--\$1.57. These results show that repository evidence can support both access policy synthesis and independent validation for applications, even when no reliable reference policy exists.

\end{abstract}

\maketitle

\vspace{-.75em}
\section{Introduction}

\input{Content/introduction}

\vspace{-.75em}
\input{Content/motivation}

\vspace{-.75em}
\input{Content/problem}

\vspace{-.75em}
\section{Dataset Construction}
\input{Content/Dataset_design}

\vspace{-.75em}
\section{Methodology}
\input{Content/methodology}

\vspace{-.75em}
\section{Experimental Design}

\input{Content/experiment_design}

\vspace{-.75em}
\section{Evaluation}
\input{Content/Eval}

\vspace{-.75em}
\section{Related Work}

\input{Content/background}

\vspace{-.75em}

\input{Content/discussion}

\vspace{-.75em}
\section{Conclusion}

\input{Content/conclusion}



\bibliographystyle{ACM-Reference-Format}
\bibliography{ref}




\end{document}

%% file: macros.tex
\usepackage{tikz}
\usetikzlibrary{
  shapes.geometric,
  arrows.meta,
  positioning
}
\usepackage{dsfont}
\usepackage{booktabs}
\usepackage{booktabs,graphicx,xcolor,colortbl}
\usepackage{multirow}
\usepackage{enumitem}
\usepackage{makecell}
\usepackage[table]{xcolor}
\usepackage{booktabs}
\usepackage{graphicx}
\usepackage{xcolor}
\usepackage{longtable}
\usepackage{comment}
\PassOptionsToPackage{hyphens}{url}
\usepackage{hyperref}
\usepackage{booktabs}
\usepackage{float}
\usepackage{amsmath}
\usepackage{algorithmic}
\usepackage{graphicx}
\usepackage{textcomp}
\usepackage{xcolor}
\usepackage{placeins}
\usepackage{caption}
\usepackage{booktabs}
\usepackage[table]{xcolor}
\usepackage{makecell}
\newcommand{\shortsectionBf}[1]{\vspace{2pt}
\noindent {\bf #1}}
\usepackage{multirow}
\usepackage{tabularx}
\usepackage{array}
\usepackage{xurl}
\usepackage{hyperref}
\usepackage{listings}
\usepackage{amsmath} \usepackage{booktabs}  \usepackage{tabularx}  \usepackage{array}
\usepackage[ruled,vlined]{algorithm2e}
\definecolor{lightgraybox}{gray}{0.94}
\newboolean{commentsOn}
\setboolean{commentsOn}{true}
\usepackage{subcaption}
\usepackage{tabularx}
\usepackage{enumitem}
\usepackage{capt-of}
\usepackage{wrapfig} 
\usepackage{tcolorbox}
\usepackage{xcolor}
\usepackage{needspace}
\newtcolorbox{findingbox}{
    colback=gray!5,
    colframe=gray!50,
    boxrule=0.5pt,
    arc=2pt,
    left=6pt,
    right=6pt,
    top=6pt,
    bottom=6pt
}

\ifthenelse{\boolean{commentsOn}}{
\newcommand{\Juan}[1]{{\color{blue}{\bf Juan:} #1}}}
{\newcommand{\Juan}[1]{}}

\ifthenelse{\boolean{commentsOn}}{
\newcommand{\pd}[1]{{\color{red}{\bf Pubali:} #1}}}
{\newcommand{\pd}[1]{}}

\usepackage{xcolor}
\usepackage{mdframed}
\newmdenv[backgroundcolor=gray!5,
    linecolor=gray,
    outerlinewidth=1pt,
    roundcorner=5,
    skipabove=2pt,
    skipbelow=0pt,
    innerleftmargin=3pt,
    innerrightmargin=3pt,
    innerbottommargin=3pt,
    innertopmargin=3pt,
    font=\normalsize
]{shadedboxed}

\ifthenelse{\boolean{commentsOn}}{
\newcommand{\pubali}[1]{{\color{red}{\bf Pubali:} #1}}}
{\newcommand{\pubali}[1]{}}
\def\BibTeX{{\rm B\kern-.05em{\sc i\kern-.025em b}\kern-.08em
    T\kern-.1667em\lower.7ex\hbox{E}\kern-.125emX}}

\ifthenelse{\boolean{commentsOn}}{
\newcommand{\ayesha}[1]{{\color{purple}{\bf Ayesha:} #1}}}
{\newcommand{\ayesha}[1]{}}

\ifthenelse{\boolean{commentsOn}}{
\usepackage{soul}

\newcommand{\sam}[1]{{\color{teal}{\bf Sam:} #1}}}
{

\newcommand{\sam}[1]{}
}

\ifthenelse{\boolean{commentsOn}}{
\newcommand{\gehao}[1]{{\color{orange}{\bf Gehao:} #1}}}
{\newcommand{\gehao}[1]{}}

\newcommand{\Sys}{{\scshape Reparo}\xspace}

%% file: Content/introduction.tex
Modern cloud platforms provide specialized policy languages for defining fine-grained access-control rules~\cite{pang2019zanzibar,cutler2024cedar,paul2024amazon}. Writing correct policies requires expertise in the language's syntax and authorization model, as well as an accurate understanding of often complex and incompletely documented authorization requirements. Consequently, manual policy writing is tedious and error-prone~\cite{yeboahduako2026permissionsloosemeasuringoverprivilege,polinsky2024}, and even small mistakes can have serious consequences~\cite{polp75}. Overly permissive policies may expose sensitive resources; for example, in 2023, Microsoft researchers accidentally exposed 38~TB of sensitive data by attaching an overly permissive token to a download link, granting full control over an internal storage account instead of a single dataset~\cite{page2023microsoft}. Overly restrictive policies, conversely, may deny legitimate users, disrupt functionality, and waste users' time; in safety-critical domains such as health care, erroneous denials can have serious consequences~\cite{bauer2011detecting}.

Policy mining and reverse-engineering techniques attempts to solve this problem through inferring authorization rules
from structured data, such as user--permission assignments~\cite{xu2014mining}, access logs~\cite{cotrini2018mining}, and relationship graphs~\cite{bui2019greedy,bui2020decision}, or from the application itself by analyzing authorization checks in code~\cite{gorski2019acminer} or observing runtime behavior~\cite{le2022automated}. However, real-world repositories rarely provide such structured input:
authorization information is spread across source code, comments,
documentation, and policy files.
Large language models (LLMs) make it feasible to integrate such scattered
evidence, but their output may contain incorrect rules and must be validated.
Existing LLM-based approaches leave three gaps. First, LLM-based policy synthesis
translates an explicit specification, such as natural-language requirements
or allowed and denied requests, into a
policy~\cite{lawal2024translating,vatsa2025synthesizing,mai2025llm,sonune2025lmn,sonune2026laps,paratore2025plain,jayasundara2026agentvlm,vatsa2026autocedar,gupta2026prose2policy,cheng2025say}.
Writing a complete specification reintroduces the manual effort that
automation seeks to remove. Second, LLM-based mining still requires
structured authorization data, such as attributes, granted permissions, or
access logs~\cite{bui2026mining,babasaheb2025towards}. Third, existing
approaches validate against a trusted reference: an input
specification~\cite{vatsa2025synthesizing}, a ground-truth
policy~\cite{bui2026mining,babasaheb2025towards}, or a human-reviewed
policy~\cite{vatsa2026autocedar}. Real repositories rarely offer such a
reference. The developer-written policy available in the repository, which we call the
\emph{repository policy}, cannot serve as one, because policy errors are
common~\cite{bauer2011detecting,yeboahduako2026permissionsloosemeasuringoverprivilege,polinsky2024}.


We therefore ask two questions. \emph{Can an executable policy that captures
an application's authorization intent be synthesized from its repository
alone? And can that policy be validated without a trusted reference?} The
repository policy, which often contains errors, is used neither as input to synthesis nor as ground truth
for validation. Our key insight is that the repository evidence used to
synthesize a policy can, independently, also serve as the reference for
validating it. Achieving this raises three challenges.


\textbf{Challenge 1: Locating and reconciling scattered authorization logic.} Authorization intent is rarely captured in a single artifact. It is scattered
across source code, comments, and documentation, most of which concerns
other behavior. Comments make up more than 20\% of non-blank code lines in popular
open-source repositories~\cite{he2019understanding}, documentation appears in
98\% of sampled GitHub repositories~\cite{pfeiffer2020constitutes}, and
authorization checks are sparse within millions of lines of
code~\cite{gorski2019acminer}. The challenge is to find the relevant
fragments and reconcile incomplete or inconsistent information during policy synthesis.

\textbf{Challenge 2: Producing valid executable policies across languages.} Policy languages differ in both syntax and authorization semantics: Cedar uses entity schemas to express role, relation, and attribute-based rules~\cite{cutler2024cedar}, OpenFGA derives access from relations between objects defined in an authorization model~\cite{aguiar2022openfga}, and Rego computes decisions through rules over structured input~\cite{opa_policy_language}. Thus, the same authorization requirement can take structurally different forms across languages. A synthesized policy must not only capture the intended authorization behavior but also pass the language's native validator~\cite{cedarValidation,opaCLI,openfgaCLI}. The challenge is to synthesize executable policies and automatically correct the errors to  enable large-scale policy synthesis.

\textbf{Challenge 3: Building a trustworthy oracle from the same evidence.}
A syntactically valid policy may still make incorrect decisions, so its behavior must be checked against an oracle, i.e., test cases pairing access requests with expected decisions. Such an oracle is rarely available in real repositories, and manually constructing one from scattered authorization evidence reintroduces the effort that automation seeks to eliminate. We therefore generate the oracle automatically using an LLM. However, LLM-generated test cases may be incorrect, ambiguous, unsupported by the evidence, or leave parts of the policy unexercised. Moreover, because the oracle and policy use the same evidence, their errors may be correlated, so agreement alone does not establish correctness. The oracle's correctness and coverage must be measured.

To address these challenges, we present \Sys, an LLM-aided pipeline that
synthesizes, repairs, and validates access-control policies from repository
evidence. \Sys decouples synthesis from validation: one set of LLMs
synthesizes the policy, and a different set derives the oracle used to
validate it. For Challenge~1, \Sys retrieves relevant content from source
code, comments, and documentation, and never reveals the repository policy to
the synthesis models. For Challenge~2, it guides generation with the target
language's grammar and examples, and iteratively repairs invalid policies
using error messages from the language's validator. For Challenge~3, the
oracle models derive test cases from the same evidence, each pairing a
concrete authorization request with an expected allow or deny decision. \Sys
runs the synthesized policy on these test cases with the language's policy engine and
compares its decisions with the expected ones. We manually assess the LLM-generated oracle: a statistically sized sample of 390 test cases is reviewed for correctness. Moreover, we measure the extent to which the test cases exercise each policy.

To evaluate \Sys, we need real application repositories that express authorization intent. Existing datasets are often synthetic~\cite{wessner2026natural,jayasundara2024ragent}, curated~\cite{bui2025abac,jayasundara2024ragent}, or small~\cite{jayasundara2024ragent}.
We therefore construct a benchmark of 228 real-world GitHub repositories that
use Cedar~\cite{cutler2024cedar}, OpenFGA~\cite{aguiar2022openfga}, or Rego~\cite{opa_policy_language}, three policy languages with distinct
authorization models. In this benchmark, 40.6--55.7\% of the repository
policies that were originally included by the developers fail their language's validator~\cite{cedarValidation,opaCLI,openfgaCLI}, confirming that they cannot
reliably serve as ground truth.

\Sys produces 99.0--100\% valid policies after repair in all three
languages. We measure correctness as \emph{decision agreement}: the share of
oracle test cases on which the policy returns the expected Allow or Deny
decision. Decision agreement is 98.2--99.6\% for OpenFGA, 82.0--86.5\% for
Cedar, and 76.3--80.2\% for Rego. Manual review finds 91.5--96.9\% of 390
sampled test cases correct. Our case studies further show that, when
repositories contain sufficient authorization evidence, synthesized policies
recover core authorization behavior and correct defects in the repository
policy.
In summary, we make the following contributions:
\begin{itemize}[leftmargin=3mm]

\item \textbf{Policy synthesis from repository evidence.} We present \Sys, an
LLM-aided pipeline that synthesizes access-control policies from code,
comments, and documentation, without a separate specification, and repairs
invalid policies using feedback from each language's validator.

\item \textbf{Validation without ground truth.} We introduce a validation
approach in which independent LLMs derive behavioral test cases from the same
repository evidence, and the language's policy engine checks the policy's
decisions against them. We also measure the correctness and coverage of these
test cases.

\item \textbf{Benchmark and evaluation.} We construct a benchmark of 228
real-world repositories across Cedar, OpenFGA, and Rego, in which
40.6--55.7\% of repository policies fail validation. Evaluated with
proprietary and open-weight models, \Sys produces 99--100\% valid policies,
and its synthesized policies can correct defects in repository policies.

\end{itemize}

%% file: Content/motivation.tex
\section{Background \& Motivation}
\label{sec:motivation}

\shortsectionBf{Access-control Policy \& Language.} An access-control policy determines whether a \emph{principal} may perform an \emph{action} on a \emph{resource} under a given \emph{context}. Permissions may be expressed through roles (RBAC), attributes (ABAC), or relationships such as ownership or membership (ReBAC). Modern applications increasingly implement such rules as \emph{policy as code}, using dedicated policy languages. Cedar~\cite{cutler2024cedar} defines rules over principals, actions, and resources and provides a schema specifying the entities used by policies. OpenFGA~\cite{aguiar2022openfga} and SpiceDB~\cite{spicedb} derive access from relationships between users and objects~\cite{pang2019zanzibar}, while Rego~\cite{opa_policy_language} uses logic rules over structured input. Unlike Cedar, OpenFGA and Rego do not provide an equivalent schema artifact. These differences make policy synthesis language-dependent, as the same authorization intent must be expressed and evaluated differently across languages. Native validators can detect syntactic and model-level errors~\cite{cedarValidation,opaCLI,openfgaCLI}.

\shortsectionBf{Test oracles.} A \emph{test oracle} specifies the expected outcome against which observed behavior is compared~\cite{barr2014oracle}. For access-control policies, a test case pairs an authorization request with an expected \textsc{Allow} or \textsc{Deny} decision~\cite{martin2006defining,xu2012model}. Constructing such an oracle typically requires a specification of intended behavior or a trusted reference policy. Without either, expected decisions must be derived from other sources. LLMs have recently been used to generate tests from software artifacts~\cite{dakhel2024effective,tang2024chatgpt}, but generated oracles still require empirical assessment.



\begin{wrapfigure}{r}{0.48\columnwidth}
\vspace{-8pt}
\centering
\includegraphics[width=\linewidth]{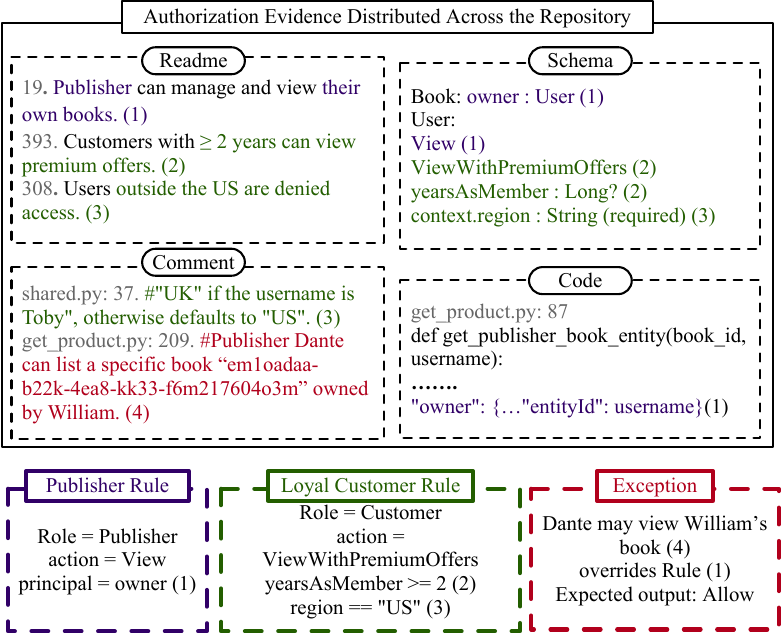}
\caption{Authorization evidence across repository artifacts and
access-control rules. Numbered fragments combine into three rules: Publisher (blue), Loyal-customer (green), and an Exception (red).}
\label{fig:motivation}
\vspace{-8pt}
\end{wrapfigure}

\shortsectionBf{Motivating example.}
Figure~\ref{fig:motivation} shows a bookstore application from a real-world repository in our dataset. No single artifact specifies its full access policy: the README describes some requirements in prose, the schema defines entity types, attributes, and actions, comments record exceptions and explain derived values, and the code shows how the application constructs the entities checked by the policy. Thus, the authorization intent is distributed across multiple artifacts. We illustrate this with three aspects of the application's authorization behavior: a publisher rule, a loyal-customer rule, and an exception to the publisher rule.

\textit{Publisher rule.} The README (line~19) states that publishers can view
their own books, but does not define what ``own'' means. The schema declares
the \texttt{View} action and gives each \texttt{Book} an \texttt{owner}
attribute of type \texttt{User}. The code in \texttt{get\_product.py}
(line~87) sets a book's \texttt{owner} to the publisher's username. Only
together do these fragments yield the rule: a publisher may view a book when
\texttt{principal == resource.owner}.

\textit{Loyal-customer rule.} The README states that customers with at least
two years of membership can view premium offers (line~393), and that users
outside the US are denied access (line~308). A comment in \texttt{shared.py}
(line~37) explains how the region is set: \texttt{"UK"} for the user Toby, and
\texttt{"US"} otherwise. The schema shows that \texttt{yearsAsMember} is
optional, while \texttt{context.region} is required. A correct policy must
combine both README requirements and check that \texttt{yearsAsMember} is
present before comparing it. Without this check, the policy fails validation.

\textit{Exception.} A comment in \texttt{get\_product.py} (line~209) states
that the publisher Dante may access a specific book owned by William. This
exception overrides the publisher rule. 


This motivating example shows the three problems that \Sys must solve. First, the
evidence for each rule is spread across four kinds of artifacts, and each
fragment is incomplete on its own. The relevant fragments must be found among
unrelated code and combined (Challenge~1). Second, the combined rule must be
written as a valid policy, which here requires details such as the presence
check on an optional attribute (Challenge~2). Third, there is no reference to
check the resulting policy against. The expected decision for a request such
as ``may Dante view William's book?'' must itself be derived from the same
fragmented evidence (Challenge~3).

%% file: Content/problem.tex
\section{Problem Statement}
\label{sec:problem}

An authorization request is a tuple $\pi=\langle p,a,r,c\rangle$ of a
principal $p$, action $a$, resource $r$, and context $c$. Let $\Pi$ be the
set of well-formed requests. A policy $P:\Pi\rightarrow
\{\textsc{Allow},\textsc{Deny}\}$ maps each request to an access decision.
We use $P_L$ to denote a policy synthesized in language $L$.

\Sys aims to synthesize a policy $P_L$ that (i) is valid in language $L$ and
(ii) is consistent with the application's intended authorization behavior.
Let $\mathcal{R}$ be an application repository and
$\mathcal{C}\subseteq\mathcal{R}$ the evidence used for policy synthesis,
including source code, comments, documentation, and, for Cedar, the schema
configuration file. Let $\mathcal{P}\subseteq\mathcal{R}$ denote the
developer-written policy files, which we call the \emph{repository policy}.
We exclude $\mathcal{P}$ from $\mathcal{C}$ during policy generation.

We check property (i) using each language's official validator, which checks syntactic validity as
well as language-specific structural, type, or compilation constraints. However, we cannot directly
measure property (ii) because the intended authorization behavior is not
explicitly available, and the repository policy $\mathcal{P}$ may contain
errors. Instead, \Sys builds a behavioral oracle $T_L$ from $\mathcal{C}$
using models different from those used for policy synthesis. $T_L$ consists
of test requests paired with expected decisions $T_L(\pi)$. Semantic validation compares $P_L(\pi)$ with $T_L(\pi)$ for each test
request, and is measured by \emph{decision agreement}. Because $T_L$ only
approximates the intended authorization behavior, its quality is
evaluated separately (RQ2).

%% file: Content/Dataset_design.tex
\label{sec:dataset-design}
Since no existing dataset provides the needed authorization evidence, \Sys
constructs a dataset of real-world GitHub repositories using Cedar, OpenFGA,
or Rego for access-control decisions, collected as of June~2026, to evaluate
policy generation from repository-level authorization evidence.
This proceeds in two stages:
(1)~repository selection, in which candidate repositories are discovered
and filtered, and (2)~extraction of the evidence $\mathcal{C}$ from each
selected repository. 

\shortsectionBf{Repository Selection.} Repositories are identified on GitHub using language-specific files: \texttt{.cedar} for Cedar; \texttt{.fga}, \texttt{.openfga}, and OpenFGA JSON model files for OpenFGA; and \texttt{.rego} for Rego. These files constitute each repository's policy $\mathcal{P}$. We count policy rules, the smallest units of $\mathcal{P}$. A Cedar \texttt{permit} or \texttt{forbid} statement, OpenFGA relation definition, and top-level Rego allow or deny rule, including helper rules, counts as one rule. We require at least five policy rules to ensure substantive use of the policy language, and $\text{stars}+\text{forks}\geq5$ at collection time to indicate minimum public interest and reduce toy or example repositories.

\shortsectionBf{Authorization-Relevant Context Extraction.}
\Sys retrieves scattered authorization evidence using BM25~\cite{robertson2009probabilistic}, which ranks repository content against queries representing authorization concepts. We define \emph{retrieval files} $\mathcal{S}\subseteq\mathcal{R}$ whose text provides the application's authorization vocabulary, including entity types, relations, and actions. 
\label{sec:extraction}

\begin{wrapfigure}[30]{r}{0.45\columnwidth}
\vspace{-1em}
\begin{minipage}{\linewidth}
\begin{algorithm}[H]
\scriptsize
\SetAlgoNlRelativeSize{-1}
\SetAlFnt{\scriptsize}
\SetAlCapFnt{\scriptsize}
\SetAlCapNameFnt{\scriptsize}
\SetKwFunction{Score}{Score}
\SetKwProg{Fn}{Function}{:}{}
\caption{Authorization-relevant context extraction}
\label{alg:context-pipeline}
\LinesNumbered
\KwIn{Repository $\mathcal{R}$; retrieval files $\mathcal{S}$; repository policy $\mathcal{P}$; threshold $\tau$}
\KwOut{Relevant context $\mathcal{C}$}

$\mathcal{C}_{\mathrm{raw}} \gets \emptyset$\;
\For{$t \in \{\textsc{code},\textsc{readme},\textsc{comment}\}$}{
  $\mathcal{C}_{\mathrm{raw}} \gets \mathcal{C}_{\mathrm{raw}} \cup \textsc{Segment}(\mathcal{R},t)$\;
}
\ForEach{$c \in \mathcal{C}_{\mathrm{raw}}$}{
  $\mathrm{score}(c) \gets \Score{$c,\mathcal{C}_{\mathrm{raw}}$}$\;
}
$\mathcal{C}' \gets \textsc{Dedup}(\mathcal{C}_{\mathrm{raw}})$\;
$\mathcal{C}' \gets \textsc{RemovePolicyLeaks}(\mathcal{C}',\mathcal{P})$\;
\ForEach{$c \in \mathcal{C}'$}{
  $\mathrm{score}(c) \gets \Score{$c,\mathcal{C}'$}$\;
}
$\mathcal{C} \gets \{c \in \mathcal{C}' : \mathrm{score}(c) > \tau\}$\;
\Return{$\mathcal{C}$}\;

\BlankLine
\Fn{\Score{$c,\mathcal{D}$}}{
\Return{$\displaystyle
\max_{s_j \in \mathcal{S}}
\left(
\mathrm{BM25}_{\mathcal{D}}(c,s_j)
+ \beta_{t(c)}\,\mathrm{BM25}_{\mathcal{D}}(\mathrm{Ctx}(c),s_j)
\right)$}\;
}
\end{algorithm}
\end{minipage}
\vspace{-1em}
\end{wrapfigure}

 
 For Cedar, $\mathcal{S}$ consists of entity schema files defining principals, resources, and actions. Because OpenFGA and Rego lack separate schema files, we use the repository policy files instead, i.e., $\mathcal{S}=\mathcal{P}$. Even invalid policies provide useful retrieval terms through application-specific identifiers. For Cedar, we identify schema files by the \texttt{.cedarschema},
\texttt{.cedarschema.json}, or \texttt{.json} extensions with `schema' in the
filename. JSON schemas are converted to Cedar's schema format using the Cedar
CLI; files that fail conversion are discarded.

Algorithm~\ref{alg:context-pipeline} shows the context extraction process. Source code, README documentation, and comments are split along their natural structure rather than into fixed-size windows (Lines~1--3): code into
top-level functions and classes via language-aware AST parsing, documentation
by section headings, and comments and docstrings kept with nearby code when
needed to preserve their meaning. 

\WFclear

Chunks are ranked with the BM25 ranking algorithm. $\textsc{Score}(c,\mathcal{D})$
(Lines~12--13) scores chunk $c$ relative to a chunk set $\mathcal{D}$, using
the tokenized full text of each file $s_j\in\mathcal{S}$ as a query. The score combines the BM25 score of a chunk with that of its surrounding
context. Each chunk takes its highest score across retrieval files, since a
chunk may be relevant to one retrieval file but not to others. All chunks are first scored relative to $\mathcal{C}_{\mathrm{raw}}$
(Lines~4--5). 
Duplicate chunks are removed (Line~6), as are chunks that have content
of $\mathcal{P}$ (Line~7). Because BM25 term weights depend on the chunk set, removing these chunks makes the
initial scores stale. So, the remaining chunks are rescored relative to
$\mathcal{C}'$ (Lines~8--9), and those scoring above $\tau$ form the final
context $\mathcal{C}$ (Line~10).


\shortsectionBf{Dataset Characterization.}
The resulting dataset contains 228 repositories: 53 for Cedar, 75 for
OpenFGA, and 100 for Rego. As indicators of practical deployment, we detect
each repository's cloud provider and containerization with a rule-based
scanner. For each repository, the README and deployment-related files are examined, prioritizing CI workflows,
infrastructure-as-code templates (e.g., Terraform, CDK), platform manifests
(e.g., \texttt{wrangler.toml}, \texttt{vercel.json}), and container files.
Files are matched against weighted provider-specific patterns (e.g., \texttt{arn:aws:} for AWS), and each repository is assigned the highest-scoring provider. A provider is identified for 124
repositories: AWS (74), GCP (17), Azure (11), and 22 across Cloudflare, Vercel, Netlify, Heroku, DigitalOcean, and Fly.io. Containerization is common, covering 50 of 74 AWS and 12 of 17
GCP repositories, and 42 of the 104 repositories without a provider. In
total, 166 of 228 repositories (72.8\%) show concrete deployment evidence.

%% file: Content/methodology.tex
\label{sec:methodology}


\begin{wrapfigure}{r}{0.45\columnwidth}
    \centering
    \vspace{-4em}
    \includegraphics[width=\linewidth]{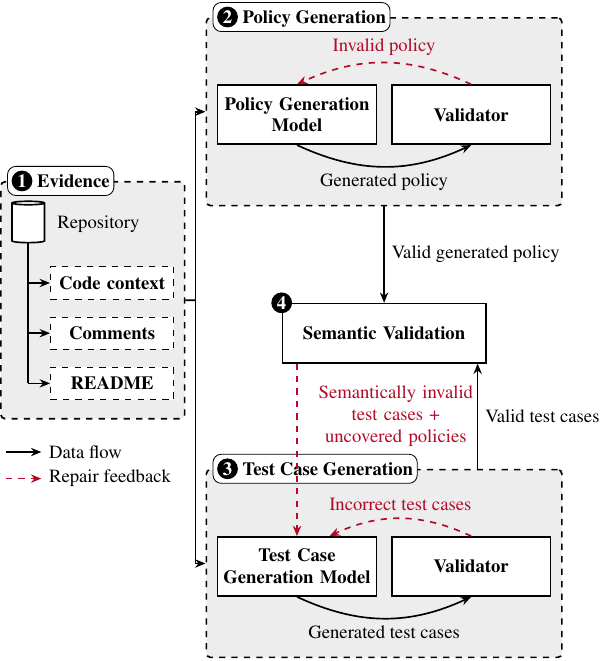}
    \caption{\Sys overview: (1) evidence extraction; (2) policy generation with validator-guided repair; (3) independent test-case generation; (4) semantic validation with feedback to refine test cases; Dashed red arrows denote repair feedback.}
    \label{fig:methodology_pipeline}
    \vspace{-1em}
\end{wrapfigure}

Section~\ref{sec:dataset-design} extracts the
authorization-relevant evidence $\mathcal{C}$ for each repository: code, comments, and
documentation ranked by Algorithm~\ref{alg:context-pipeline}. This section describes
how \Sys uses that evidence to synthesize an access-control policy
and to validate it, without requiring a separate specification.

Figure~\ref{fig:methodology_pipeline} summarizes the workflow. Starting from
$\mathcal{C}$, \Sys runs two separate tracks. The \emph{synthesis}
track (Section~\ref{sec:policy-synthesis}) generates a policy with an LLM,
assesses its policy validity using the language's validation tool, and repairs
any reported errors. The \emph{validation} track
(Section~\ref{sec:semantic-assessment}) uses a different model to derive
behavioral test cases from the same evidence. \Sys then evaluates the
synthesized policy on these test cases using the language's policy engine and
measures decision agreement with the expected outcomes.

The validation result is useful beyond a single agreement score. Each mismatch
identifies a concrete request on which the policy's decision differs from the
decision the evidence supports, together with the evidence lines behind the
expected decision. A developer can use these mismatches to locate and edit a
faulty rule, or to discard a test case that misread the repository evidence $\mathcal{C}$. We use the bookstore repository from
Section~\ref{sec:motivation} as a running example throughout this section.

\subsection{Policy Synthesis \& Repair}
\label{sec:policy-synthesis}

\begin{wrapfigure}{r}{0.5\columnwidth}
    \centering
    \vspace{-0.8em}
    \includegraphics[width=\linewidth]{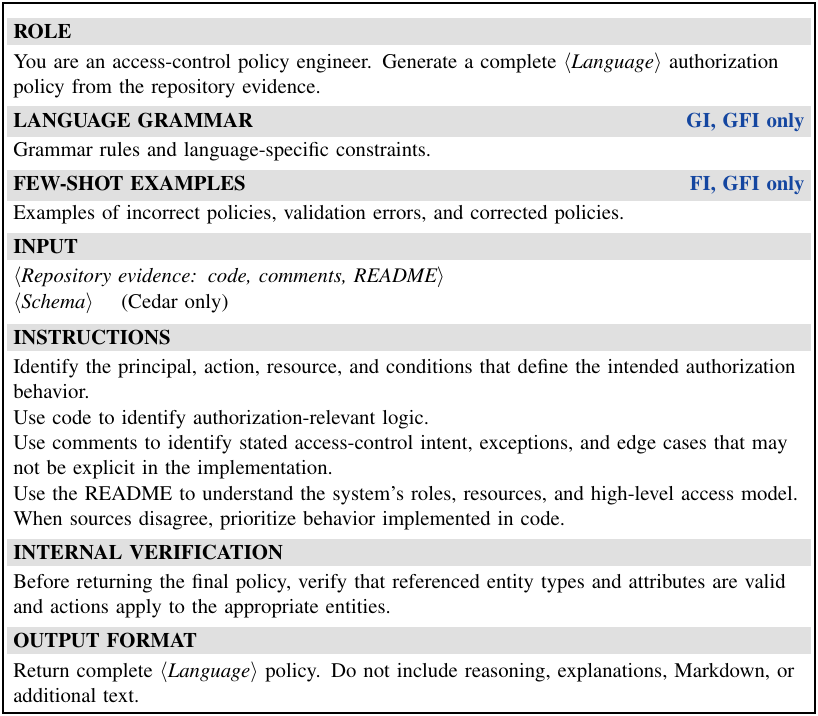}
    \caption{High-level structure of the policy-generation prompt.}
    \label{fig:prompt_gen}
    \vspace{-.9em}
\end{wrapfigure}
 
\subsubsection{Policy Synthesis}
\label{sec:policy-generation}
An LLM recovers the repository's implicit authorization intent as a
policy $P_L$, where $L\in\{\text{Cedar},\text{OpenFGA},\text{Rego}\}$. The
model receives the repository evidence $\mathcal{C}$ together with language-specific
guidance; for Cedar, the schema is also provided during generation because it defines
the entity types and attributes required to validate $P_L$. The repository
policy $\mathcal{P}$ is never provided.

\textit{Prompt variants.}
We attempted four prompt variants for each language for this synthesis  task. The baseline, \texttt{I},
contains the task description, the repository evidence $\mathcal{C}$, extraction
instructions, a self-verification step, and the output format. \texttt{GI}
adds the syntax and key rules of the target language, taken from the official
Cedar~\cite{cedar_policy_syntax_grammar},
OpenFGA~\cite{openfga_language_grammar}, and
Rego~\cite{opa_policy_reference_grammar,opa_rego_bnf} grammars. \texttt{FI}
adds few-shot examples, and \texttt{GFI} combines the grammar and few-shot examples. Each few-shot example
shows an incorrect policy, its validator error logs,
and the corrected policy. The examples cover most frequent policy generation errors,
including incorrect attribute or relation references, invalid expressions,
unsafe variables, deprecated features, and violations of the schema or
authorization model.

\begin{wrapfigure}{r}{0.55\columnwidth}
    \centering
    \vspace{-0.8em}
    \includegraphics[width=\linewidth]{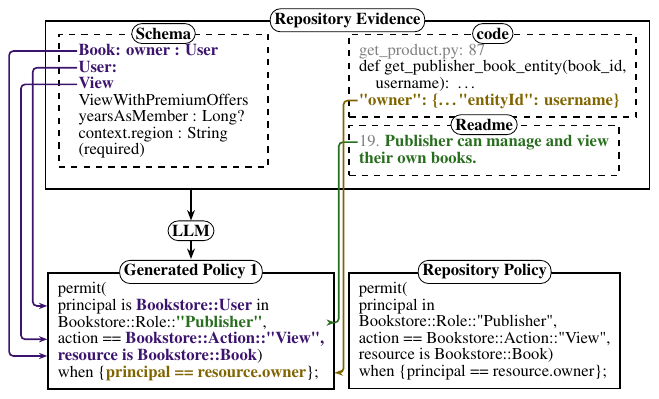}
    \caption{The \textit{Publisher rule} synthesized from schema, code, and README evidence; arrows map each fragment to the policy element it determines. The repository policy is shown for reference.}
    \label{fig:gen_example}
    \vspace{-1em}
\end{wrapfigure}

The model is instructed (Figure \ref{fig:prompt_gen}) to identify the principal, action, resource, and
conditions that define each authorization behavior. Code is used to identify
authorization logic, such as role and permission checks, ownership; comments capture intent,
exceptions, and edge cases not evident from the code; and the README conveys
the system's roles, resources, and high-level access model. When sources show conflicting evidence, the source code behavior is prioritized. These examples are generic and do not leak information from the repository being evaluated. Our evaluation shows that the \texttt{GFI} prompt improves policy generation
the most, while the \texttt{I} prompt produces substantially more errors (see Section~\ref{sec:ablations}).


Figure~\ref{fig:gen_example} shows the policy synthesis of the \textit{Publisher rule}. The README states that publishers can view their own published books, while the code represents ownership through the \texttt{owner} attribute of a \texttt{Book}; the generated policy combines these two pieces of evidence into the \textit{Publisher rule} of Section \ref{sec:motivation}. We show the repository policy $\mathcal{P}$, which is valid and correct for this repository, as a reference to demonstrate that the generated policy captures the intended authorization behavior.

\subsubsection{Policy Validation and Repair}
\label{sec:policy-repair}
 
A policy $P_L$ is \emph{valid} (objective~(i), Section~\ref{sec:problem})
if it passes the language-specific validator for $L$ without errors: the Cedar policy validator, \texttt{opa check} for Rego, and
\texttt{fga model validate} for
OpenFGA~\cite{cedarValidation,opaCLI,openfgaCLI}. These validators check
syntactic validity as well as language-specific structural, type, or
compilation constraints, but do not establish that the policy captures the
repository's intended authorization behavior. When $P_L$ fails
validation, an LLM repairs it, modifying only what is necessary to resolve
the errors. The repaired
$P_L$ is checked again, and the loop continues until $P_L$ is valid or the
attempt limit is reached. The limit is two attempts for OpenFGA and
Rego, whose outputs are validated as a single authorization model, and four
for Cedar, whose policies contain multiple statements that can fail
independently.

\begin{wrapfigure}{r}{0.55\columnwidth}
    \centering
    \vspace{-0.8em}
    \includegraphics[width=\linewidth]{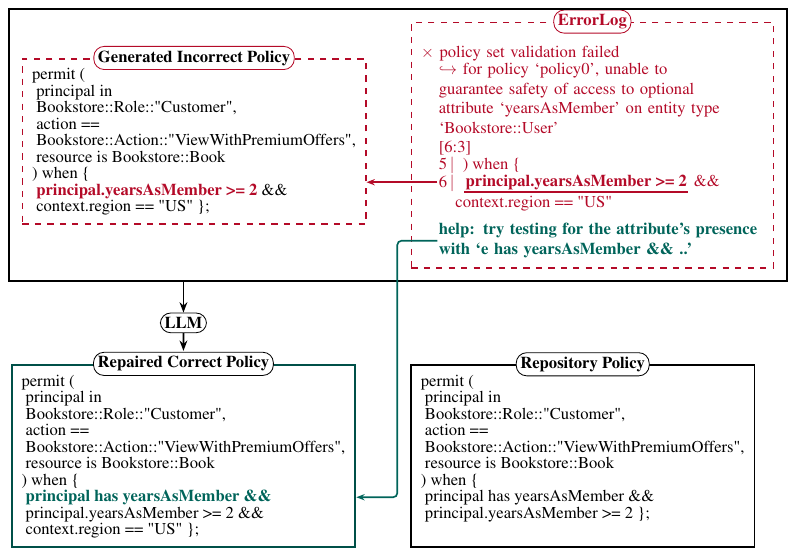}
    \caption{Error-log-guided repair: the validator error (red) flags unsafe access to an optional attribute, and the LLM adds the required fix (green). The repository policy is shown for reference.}
    \vspace{-.8em}
    \label{fig:repair_example}
\end{wrapfigure}

\textit{Prompt variants.}
Three repair variants differ only in the supporting information provided. In
\texttt{EL} (\emph{error log}), the model receives the invalid policy and its
validator error log. In \texttt{FS-C} (\emph{few-shot correct}), it receives the invalid policy and three structurally similar valid policies from other repositories, retrieved using BM25. We exclude the target repository's own policy $\mathcal{P}$ so that the LLM
cannot simply copy the target policy, particularly because repository policies
may themselves contain errors. In \texttt{FS-R} (\emph{few-shot repair}), it
receives the invalid policy and three similar repair examples, each consisting
of an invalid policy, its validator error log, and the corrected policy. All
variants share the same instructions and output format. Among these variants, \texttt{EL} performs best and is used in the main repair experiments (see Section~\ref{sec:ablations}).
Figure~\ref{fig:repair_example} illustrates an \texttt{EL} repair of the
\textit{Loyal-customer rule} of Section~\ref{sec:motivation}. The generated policy incorrectly accesses the
optional attribute \texttt{principal.yearsAsMember} without first checking
its presence, which the Cedar validator rejects as unsafe. The LLM adds
the required check. Policy validation establishes
only that $P_L$ is well-formed; whether its decisions match the repository's
intended authorization behavior (objective~(ii)) is validated in
Section~\ref{sec:semantic-assessment}.

\subsection{Semantic Validation}
\label{sec:semantic-assessment}
\Sys constructs a behavioral oracle $T_L$ from the evidence $\mathcal{C}$ used
for policy synthesis. The oracle provides validation tests that capture the
authorization behavior implied by this evidence, allowing us to assess
whether the generated policy captures the repository's intended behavior.
This section describes how \Sys constructs $T_L$ and uses it for semantic
validation.

\subsubsection{Oracle Generation}
\label{sec:oracle-generation}


For each repository, the oracle $T_L$ consists of test cases containing a
request, the entity or relationship data needed to execute it, and an
expected \textsc{Allow} or \textsc{Deny} decision. Tests are generated from
the same evidence as the policy, but using distinct model families to reduce
self-preference bias~\cite{panickssery2024llm}, where a model favors outputs
similar to its own.



\textit{Prompt variants.}
Two variants are evaluated. The baseline, \texttt{I}, contains the task
instructions, the evidence, and the output format.
\texttt{GI} adds the grammar and formatting rules of the target language, so
that requests use the representation expected by its evaluator. Between the two, \texttt{GI}
produces more executable test cases and is therefore used in the main
experiments.

\begin{wrapfigure}{r}{0.55\columnwidth}
    \centering
    \vspace{-1em}
    \includegraphics[width=\linewidth]{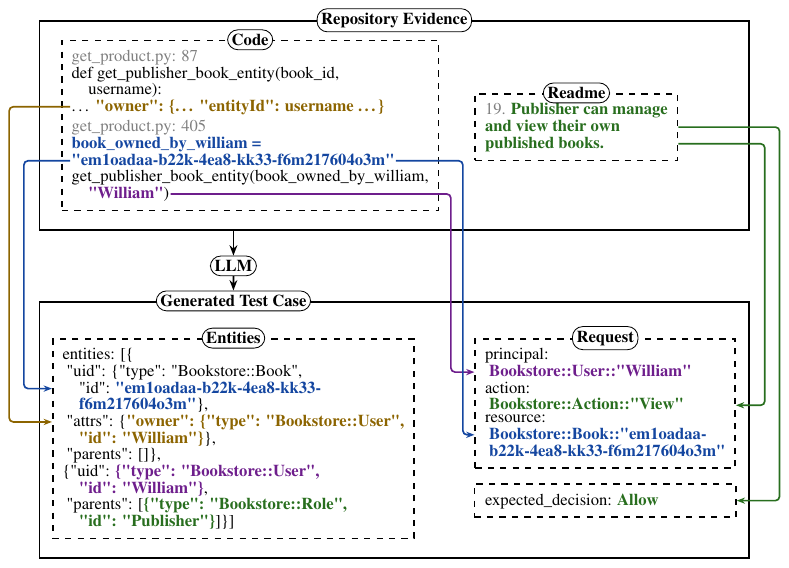}
    \caption{An oracle test case derived from code and README evidence, consisting of entity data, a request, and the expected decision (Allow); arrows trace each field to its source.}
    \label{fig:test_gen_example}
    \vspace{-1em}
\end{wrapfigure}

Since each test case must use the names the policy under test expects, the
oracle model receives only the interface of $P_L$: its types and relation names for
OpenFGA, and its input variables for Rego. For Cedar, the schema
already fixes this interface, so no information from $P_L$ is provided. For
each authorization scenario supported by $\mathcal{C}$, the model identifies
the principal $p$ (who is requesting access), action $a$ (what they want to
do), resource $r$ (what they want to access), and context $c$,
and constructs a complete test request. When the evidence supports both
outcomes, it generates paired \textsc{Allow} and \textsc{Deny} cases by
varying a single evidenced condition, principal, resource, or relationship
while keeping the rest of the scenario fixed. Each test case records a short
rationale, the line numbers of supporting evidence, and any conflicting
evidence, making every expected decision traceable to repository evidence $\mathcal{C}$.
Figure~\ref{fig:test_gen_example} continues the running example with a test case for the \textit{Publisher rule} of Section~\ref{sec:motivation}. The code at \texttt{get\_product.py:87} shows that the repository records a book's publisher in its \texttt{owner} attribute, and line~405 binds a concrete book to the user \texttt{William}. The test therefore instantiates \texttt{William} as a member of \texttt{Role::"Publisher"} and a book whose \texttt{owner} is William, and issues a request in which William views that book. Since the README permits publishers to view their own books, the expected decision is \textsc{Allow}.


\subsubsection{Semantic Validation \& Oracle Refinement}
\label{sec:oracle-repair}

\begin{wrapfigure}{r}{0.55\columnwidth}
    \centering
    \vspace{-1em}
    \includegraphics[width=\linewidth]{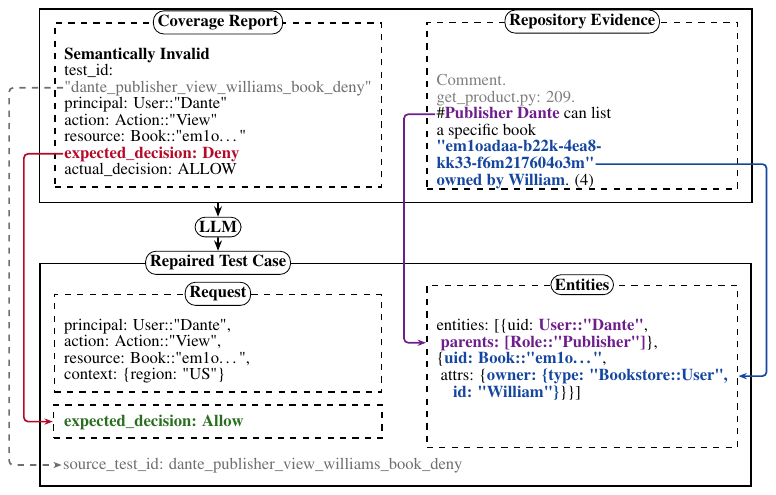}
    \caption{Test-case refinement: a case that wrongly expected Deny is re-derived from a comment and corrected to Allow.}
    \label{fig:test-repair-example}
    \vspace{-1em}
\end{wrapfigure}

Non-executable test cases are set aside, and the remaining cases
are executed against $P_L$. Semantic validation checks whether the policy
makes the same \textsc{Allow} or \textsc{Deny} decision as the oracle for each
executable test request. \Sys then refines the oracle test suite based on the
generated policy and its coverage. We compile a coverage report identifying
non-executable test cases, disagreeing test cases
with $P_L$, and policy rules that are not exercised by the
test suite. For each disagreement, \Sys initially treats the test case as semantically
invalid and re-checks it by re-deriving its expected decision from the
repository evidence. If the evidence supports the oracle decision, the
disagreement indicates incorrect policy behavior.

\paragraph{Prompt.}
The refinement prompt uses a similar \texttt{GI} prompt structure and repository
evidence as oracle generation, together with the coverage report. The report guides which cases to refine,
while expected decisions remain based solely on repository evidence
$\mathcal{C}$. The model re-derives expected decisions for flagged cases,
corrects invalid requests and their entity data, and adds cases for
uncovered rules and under-covered conditions. New test cases are added only
when supported by the evidence, and refined cases are returned in the same
structured format. Figure~\ref{fig:test-repair-example} shows a refinement from the same
bookstore repository. The original test case expected \textsc{Deny}, but the policy
returned \textsc{Allow}. Re-derivation showed that a code comment indicates
that the requesting user may access the specific book, revealing that the
original case had misinterpreted the evidence. Therefore, the expected decision is corrected to \textsc{Allow}.

%% file: Content/experiment_design.tex
\paragraph{Model Selection}

\Sys is evaluated with four contemporary models from distinct providers: Claude Sonnet~5~\cite{anthropic2026claude},
DeepSeek-V4-Pro~\cite{deepseek_v4_pro}, GPT-5.4~\cite{openai_gpt5_4}, and
GLM-5.2~\cite{glm52}. The models are accessed through their providers' APIs.

\begin{wraptable}{r}{0.38\textwidth}
\vspace{-2em}
\centering
\scriptsize
\setlength{\tabcolsep}{3pt}
\begin{tabular}{@{}lll@{}}
\toprule
Phase & Variant & Information \\
\midrule
\multirow{4}{*}{\shortstack[l]{Policy\\generation}}
  & \texttt{I}           & Evidence + instructions \\
  & \texttt{GI}          & \texttt{I} + grammar \\
  & \texttt{FI}          & \texttt{I} + few-shot examples \\
  & \texttt{GFI}$^\star$ & \texttt{I} + grammar + few-shot \\
\midrule
\multirow{3}{*}{\shortstack[l]{Policy\\repair}}
  & \texttt{EL}$^\star$  & Invalid policy + error log \\
  & \texttt{FS-C}        & Invalid policy + 3 valid policies \\
  & \texttt{FS-R}        & Invalid policy + 3 repair examples \\
\midrule
\multirow{2}{*}{\shortstack[l]{Oracle\\generation}}
  & \texttt{I}           & Evidence + instructions \\
  & \texttt{GI}$^\star$  & \texttt{I} + test-case grammar \\
\midrule
\shortstack[l]{Oracle\\refinement}
  & \texttt{GI}          & \texttt{GI} + coverage report \\
\bottomrule
\end{tabular}

\vspace{2pt}
\caption{Prompt variants per phase. $\star$ marks the best-performing
variant.}
\label{tab:prompt-variants}
\vspace{-3em}
\end{wraptable}

All four are first benchmarked on policy generation
with the \texttt{GFI} (Table~\ref{tab:prompt-variants}) 
prompt on a pilot set of 8 repositories per language.
Claude Sonnet~5 and DeepSeek-V4-Pro achieve the highest policy validity (Section~\ref{sec:policy-repair}), with 100\% of Cedar policy
statements and OpenFGA authorization models passing validation and 87.5\% of
Rego authorization models. GPT-5.4 achieves
98.3\%, 100\%, and 75.0\%, while GLM-5.2 achieves 97.9\%, 87.5\%, and 75.0\%,
for Cedar, OpenFGA, and Rego, respectively. Claude Sonnet~5 and
DeepSeek-V4-Pro are therefore used for policy generation and repair, and
GPT-5.4 and GLM-5.2 for oracle generation and refinement.



\paragraph{Retrieval Configuration}
\label{sec:retrieval-config}
Supplying the full repository is often infeasible and, even when it fits, undesirable: LLM performance degrades when relevant information is embedded in long contexts~\cite{liu2024lost}, while retrieval improves performance across context-window sizes~\cite{xu2024retrieval}, including repository-level code tasks with information scattered across files~\cite{zhang2023repocoder}. \Sys therefore supplies only the evidence $\mathcal{C}$ selected by Algorithm~\ref{alg:context-pipeline}.

Since the amount of authorization-relevant evidence varies across repositories, a fixed top-$k$ cutoff may discard useful evidence or include irrelevant content. We instead use a score threshold $\tau$, selected through a pilot study on six repositories, two per language, with $\tau\in\{0,10,20,30,40,50\}$. We manually inspect the retrieved contexts and resulting policies for each value and select $\tau=20$, which removes weakly related chunks. When the selected evidence fits within the context window, we include all of it. Otherwise, chunks are added in decreasing relevance-score order until the context window is reached.

\subsection{Design-Choice Ablations}
\label{sec:ablations}

\begin{wraptable}{r}{0.450\columnwidth}
\centering
\scriptsize
\vspace{-2em}
\setlength{\tabcolsep}{2.5pt}
\begin{tabular}{llcc}
\toprule
\textbf{Lang.} & \textbf{Variant} & \textbf{Pre-repair} & \textbf{Post-repair} \\
\midrule
\multirow{4}{*}{Cedar}
& \texttt{I}   & 74/84 (88.1\%) & 84/84 (100\%) \\
& \texttt{FI}  & 76/81 (93.8\%) & 81/81 (100\%) \\
& \texttt{GI}  & 88/89 (98.9\%) & 89/89 (100\%) \\
& \texttt{GFI} & 86/86 (100\%)  & 86/86 (100\%) \\
\midrule
\multirow{4}{*}{OpenFGA}
& \texttt{I}   & 11/15 (73.3\%) & 15/15 (100\%) \\
& \texttt{FI}  & 12/15 (80.0\%) & 15/15 (100\%) \\
& \texttt{GI}  & 13/15 (86.7\%) & 15/15 (100\%) \\
& \texttt{GFI} & 14/15 (93.3\%) & 15/15 (100\%) \\
\midrule
\multirow{4}{*}{Rego}
& \texttt{I}   & 12/20 (60.0\%) & 19/20 (95.0\%) \\
& \texttt{FI}  & 14/20 (70.0\%) & 19/20 (95.0\%) \\
& \texttt{GI}  & 11/20 (55.0\%) & 20/20 (100\%) \\
& \texttt{GFI} & 18/20 (90.0\%) & 20/20 (100\%) \\
\bottomrule
\end{tabular}
\caption{Policy validity per generation prompt, before and after error-log repair.}
\vspace{-4.5em}
\label{tab:prompt-ablation}
\end{wraptable}

The following ablations determine the configuration used in the main
experiments. They are conducted with DeepSeek-V4-Pro on 20\% of
each language's repositories: 12 Cedar, 15 OpenFGA, and 20 Rego repositories.

\paragraph{Policy Generation Prompt Variants.}
The four generation variants (\texttt{I}, \texttt{FI}, \texttt{GI},
\texttt{GFI}) are compared in Table~\ref{tab:prompt-ablation}. \texttt{GFI},
which combines grammar and few-shot guidance, achieves the highest pre-repair
policy validity in all three languages: 100\% for Cedar, 93.3\% for OpenFGA, and
90.0\% for Rego. The two components help differently across languages.
Grammar guidance (\texttt{GI}) yields the larger gain for Cedar and OpenFGA,
whereas few-shot examples (\texttt{FI}) yield the larger gain for Rego, where
grammar alone slightly reduces validity relative to \texttt{I}.
\texttt{GFI} is used in the main experiments.

\begin{findingbox}
\textbf{Finding 1.}
Grammar guidance provides the largest gains for Cedar and OpenFGA, and
few-shot examples for Rego; combining both (\texttt{GFI}) yields the most
reliable generation in all three languages and reduces the repair required.
\end{findingbox}

\paragraph{Policy Repair Prompt Variants.}

The three repair variants (\texttt{EL}, \texttt{FS-C}, \texttt{FS-R}; see Table~\ref{tab:prompt-variants}) are
applied to all invalid policies produced across the four generation variants
in Table~\ref{tab:prompt-ablation}: 16 Cedar statements, 10 OpenFGA models, and
25 Rego policies. A repair succeeds if the policy becomes valid within the attempt limit. \texttt{EL} performs best, repairing
all Cedar and OpenFGA policies and 92\% of Rego policies. The two remaining
Rego failures illustrate the limits of validator-guided repair: in one, the
policy used a deprecated language feature for which the validator suggested
no replacement; in the other, the policy required restructuring to resolve a
dependency. \texttt{FS-C} repairs none of the invalid policies. 

\begin{wraptable}{r}{0.40\columnwidth}
\centering
\scriptsize
\label{tab:policy-repair-results}
\setlength{\tabcolsep}{2.5pt}
\begin{tabular}{lccc}
\toprule
\textbf{Variant} & \textbf{Cedar} & \textbf{OpenFGA} & \textbf{Rego} \\
\midrule
\texttt{EL}   & 16/16 (100\%) & 10/10 (100\%) & 23/25 (92\%) \\
\texttt{FS-C} & 0/16 (0\%)    & 0/10 (0\%)    & 0/25 (0\%) \\
\texttt{FS-R} & 16/16 (100\%) & 9/10 (90\%)   & 18/25 (72\%) \\
\bottomrule
\end{tabular}
\caption{Share of invalid policies repaired by each repair prompt variant.}
\vspace{-2em}
\end{wraptable}

This result suggests that examples of correct policies alone provide limited guidance for repairing an invalid target policy. Without information about the error, the model may reproduce content from the retrieved examples or return the target unchanged. Correct policies therefore do not indicate what is wrong with the target. \texttt{FS-R}, which provides explicit error-to-repair examples, performs substantially better but remains below \texttt{EL}. We use \texttt{EL} in the main experiments; it also requires no additional example retrieval.

 
\begin{findingbox}
\textbf{Finding 2.}
Repair depends on error-specific guidance. Correct-policy examples alone are
ineffective, error-to-repair examples help substantially, and the target
policy's own validator feedback is the most effective solution.
\end{findingbox}

\paragraph{Oracle Generation Prompt Variants.}
The instruction-only (\texttt{I}) and grammar-augmented (\texttt{GI}) oracle
variants are compared on the same subset. \texttt{GI} produces a higher
proportion of executable test cases and is used in the main experiments.

\paragraph{Evidence sources.}
\begin{wraptable}{r}{0.50\columnwidth}
\vspace{-1em}
\centering
\scriptsize

\setlength{\tabcolsep}{2pt}
\vspace{-1em}
\begin{tabular}{lcc}
\toprule
\textbf{Setting} & \textbf{Valid} & \textbf{Decision} \\
                 & \textbf{Policy} & \textbf{Agreement} \\
\midrule
All sources (\Sys) & 86/86 & \textbf{91.9\%} \\
\midrule
\multicolumn{3}{l}{\emph{Leave one source out}} \\
\quad $-$ Code     & 64/64 & 58.4\%\\
\quad $-$ Comments &  53/53 & 55.9\% \\
\quad $-$ Docs     &  40/41 & 47.7\% \\
\midrule
\multicolumn{3}{l}{\emph{Single source}} \\
\quad Code only$^{a}$ & 47/47 & 63.8\% \\
\quad Docs only$^{b}$ & 54/55 & 57.9\%\\
\quad Comments only   & 52/52 & 52.0\%\\
\bottomrule
\multicolumn{3}{l}{\tiny $^{a}$Input setting of code-based extraction~\cite{gorski2019acminer,li2021automatic,zift};} \\
\multicolumn{3}{l}{\tiny $^{b}$input setting of NL-to-policy generation~\cite{xiao2012automated,gupta2026prose2policy,vatsa2025synthesizing}.}
\end{tabular}
\caption{Policy validity and decision agreement when evidence sources are removed or used alone (12 Cedar repositories).}
\label{tab:evidence-ablation}
\vspace{-3em}
\end{wraptable}
To assess whether combining different evidence sources is necessary, we either remove each source from $\mathcal{C}$ or use it alone. This ablation is run on the 12 Cedar repositories using DeepSeek-V4-Pro
and evaluated against a generated test oracle containing 442 test cases
derived from the same evidence. We also provide the schema during generation because it is required by the policy validation tool. Table~\ref{tab:evidence-ablation} shows that removing any single source
reduces decision agreement, because each source provides complementary
authorization information. Removing a source also reduces the number
of generated policies, from 86 with all sources to 64 without
code, 53 without comments, and 40 without documentation. The single-source settings serve as input-setting baselines: they
reproduce the inputs available to prior code-based~\cite{gorski2019acminer,li2021automatic,zift}
and NL-to-policy~\cite{xiao2012automated,gupta2026prose2policy,vatsa2025synthesizing}
approaches within the same pipeline, isolating the effect of combining
evidence sources. Decision agreement reaches 63.8\% with code only and
57.9\% with documentation only, compared with 91.9\% with all sources.





\begin{findingbox}
\textbf{Finding 3.}
Combining heterogeneous evidence sources is necessary for faithful policy generation. Removing any
single source reduces agreement, mainly by
omitting authorization rules, and the single-source settings corresponding to
code-based extraction and NL-to-policy generation reach only 63.8\% and
57.9\%, compared with 91.9\% for \Sys.
\end{findingbox}

%% file: Content/Eval.tex

We evaluate \Sys through the following research questions:

\shortsectionBf{RQ1:} To what extent can LLMs generate valid access-control policies from repository evidence $\mathcal{C}$, and how well do their authorization decisions agree with independently generated oracles $T_L$?

\shortsectionBf{RQ2:} How reliable and comprehensive are the generated behavioral test oracles $T_L$?

\shortsectionBf{RQ3:} How and why do generated policies $P_L$ differ from the repository policies $\mathcal{P}$?

We additionally report the per-repository cost of \Sys, including token consumption, API cost, and execution time for policy generation, repair, oracle generation, and refinement.

\begin{wraptable}{r}{0.5\columnwidth}
\vspace{-1em}
\centering
\scriptsize
\setlength{\tabcolsep}{3pt}
\begin{tabular}{llcc}
\toprule
\textbf{Model} & \textbf{Lang.} &
\textbf{Pre-repair (\texttt{GFI})} & \textbf{Post-repair (\texttt{EL})} \\
\midrule
\multirow{3}{*}{Claude S5}
& Cedar   & 880/887 (99.2\%) & 887/887 (100\%) \\
& OpenFGA & 73/75 (97.3\%)   & 75/75 (100\%) \\
& Rego    & 89/100 (89.0\%)  & 100/100 (100\%) \\
\midrule
\multirow{3}{*}{DeepSeek V4}
& Cedar   & 816/862 (94.7\%) & 854/862 (99.1\%) \\
& OpenFGA & 67/75 (89.3\%)   & 75/75 (100\%) \\
& Rego    & 93/100 (93.0\%)  & 99/100 (99.0\%) \\
\bottomrule
\end{tabular}
\caption{Policy validity (valid/total) on all 228 repositories, before and after repair.}
\label{tab:gfi-results}
\vspace{-3.5em}
\end{wraptable}
\shortsectionBf{RQ1- Policy Synthesis:}
\label{sec:rq1}
RQ1 assesses the generated policy $P_L$ through policy validation and
semantic validation. \emph{Policy validity} is the proportion of
generated policies accepted by the corresponding language-specific
validator. \emph{Semantic validation} uses the behavioral oracle $T_L$,
which consists of test requests paired with expected authorization
decisions: each test request is executed against $P_L$, and the returned
decision is compared with the oracle's expected decision. Semantic
validation is measured by \emph{decision agreement}, the fraction of
executable test cases for which the Allow/Deny decision returned by
$P_L$ matches the oracle's expected decision.

\begin{wraptable}{r}{0.54\columnwidth}
\vspace{-1em}
\centering
\scriptsize
\setlength{\tabcolsep}{2.8pt}
\begin{tabular}{lllr}
\toprule
\textbf{Lang.} & \textbf{Policy} & \textbf{Oracle} & \textbf{Decision Agreement} \\
\midrule
\multirow{4}{*}{Cedar}
& Claude S5   & GPT-5.4 & 86.5\% ($n$=1213) \\
& DeepSeek V4 & GPT-5.4 & 85.0\% ($n$=1130) \\
& Claude S5   & GLM-5.2 & 82.9\% ($n$=1063) \\
& DeepSeek V4 & GLM-5.2 & 82.0\% ($n$=1060) \\
\midrule
\multirow{4}{*}{OpenFGA}
& Claude S5   & GPT-5.4 & 98.3\% ($n$=4224) \\
& DeepSeek V4 & GPT-5.4 & 99.6\% ($n$=4135) \\
& Claude S5   & GLM-5.2 & 98.7\% ($n$=2811) \\
& DeepSeek V4 & GLM-5.2 & 98.2\% ($n$=2510) \\
\midrule
\multirow{4}{*}{Rego}
& Claude S5   & GPT-5.4 & 76.3\% ($n$=2473) \\
& DeepSeek V4 & GPT-5.4 & 77.1\% ($n$=1461) \\
& Claude S5   & GLM-5.2 & 78.8\% ($n$=1718) \\
& DeepSeek V4 & GLM-5.2 & 80.2\% ($n$=916) \\
\bottomrule
\end{tabular}
\caption{Decision agreement between generated policies and oracles per model pair; n is the number of executed test requests.}
\label{tab:decision-match}
\vspace{-3em}
\end{wraptable}

\paragraph{Policy Validity.}
Table~\ref{tab:gfi-results} reports policy validity before and after repair. After
repair, Claude Sonnet~5 reaches 100\% validity in all three languages, and
DeepSeek-V4-Pro reaches 99.1\% for Cedar, 100\% for OpenFGA, and 99.0\% for
Rego. Repair contributes most for the weaker starting points, e.g., raising
DeepSeek-V4-Pro from 89.3\% to 100\% on OpenFGA and Claude Sonnet~5 from
89.0\% to 100\% on Rego.

All eight invalid Cedar policy statements generated by DeepSeek-V4-Pro after repair come from a single repository and involve
language- and schema-level errors, such as comparisons between incompatible
types and invalid use of language constructs; the validator flags them, but
its error log does not indicate the correct repair. The invalid Rego
policy after the repair stage depends on TFLint-specific built-ins that the repository's repository
policy also uses; since the validator error log does not describe such
library dependencies, the repair model cannot infer the required execution
environment. 

\paragraph{Semantic validation.}

Table~\ref{tab:decision-match} reports how often the synthesized policies
agree with the oracles. Decision agreement differs little across model pairs, but
strongly across languages: 98.2--99.6\% for OpenFGA, 82.0--86.5\% for Cedar,
and 76.3--80.2\% for Rego. This pattern is explained by how many conditions a decision
depends on (see test-case complexity in RQ2). Most OpenFGA test cases depend on zero or
one condition, while more than a third of Cedar and Rego cases depend on six
or more. Thus, repair makes nearly every policy valid, but agreement still
drops as actual authorization decisions combine more conditions.

\begin{findingbox}
\textbf{RQ1 Answer.}
With validator-guided repair, \Sys produces valid policies in 99--100\% of
cases. Decision agreement is language-dependent: 98.2--99.6\% for OpenFGA,
82.0--86.5\% for Cedar, and 76.3--80.2\% for Rego, with lower agreement for
languages whose decisions depend on more conditions.
\end{findingbox}

\paragraph{Smaller open-weight models.}
\label{sec:small-model}

We additionally evaluate two smaller open weight models for policy generation and repair, Qwen3.5-27B~\cite{qwen35_27b} and
Gemma-4-12B-IT~\cite{gemma4}. Gemma-4-31B~\cite{gemma4} and Qwen3.5-9B~\cite{qwen35} were also explored
for oracle generation but produced too few usable test cases for quantitative
evaluation and are therefore excluded from the oracle analysis. These models
were hosted on a high-performance computing (HPC) cluster. Table~\ref{tab:small-model-results} reports policy validity for the two smaller
models. Repair improves policy validity for both, but the gains are smaller and less
uniform than for the larger models: Qwen3.5-27B reaches 88.1--92.1\% after
repair, while Gemma-4-12B reaches 92.2\% for Cedar but only 39.7\% for
OpenFGA and 68.7\% for Rego.

\begin{wraptable}{r}{0.45\columnwidth}
\vspace{-8pt}
\centering
\tiny
\setlength{\tabcolsep}{2.5pt}
\begin{tabular}{llcc}
\toprule
\textbf{Model} & \textbf{Lang.} & \textbf{Pre-repair} & \textbf{Post-repair} \\
\midrule
\multirow{3}{*}{Qwen3.5-27B}
& Cedar   & 128/159 (80.5\%) & 140/159 (88.1\%) \\
& OpenFGA & 30/38 (78.9\%)   & 35/38 (92.1\%) \\
& Rego    & 55/71 (77.5\%)   & 63/71 (88.7\%) \\
\midrule
\multirow{3}{*}{Gemma-4-12B}
& Cedar   & 669/1025 (65.3\%) & 945/1025 (92.2\%) \\
& OpenFGA & 19/63 (30.2\%)    & 25/63 (39.7\%) \\
& Rego    & 50/83 (60.2\%)    & 57/83 (68.7\%) \\
\bottomrule
\end{tabular}
\caption{Policy validity of smaller open-weight models, before and after repair.}
\label{tab:small-model-results}
\vspace{-3em}
\end{wraptable}

The main limitation arises before generation. Although both models advertise
long context windows, the effective input limit varied across requests in our
deployment: some requests failed at approximately 81K tokens, while others
accepted substantially larger inputs. The error did not report the cause; the most likely explanation is
that the capacity of the serving framework (SGLang~\cite{zheng2024sglang}),
configured with a static memory fraction of 0.78, depended on the GPU memory
available to each job on the HPC cluster. Prior work has further shown that the usable
context of long-context models is often substantially shorter than
advertised~\cite{hsieh2024ruler,liu2024lost}. This affects 12 Cedar, 27 OpenFGA, and 21 Rego
repositories for Qwen3.5-27B, and 14, 12, and 17 repositories for
Gemma-4-12B with no output. These cases reflect context limitations rather than generation
failures and are excluded from Table~\ref{tab:small-model-results}. We also exclude 19 refusals and 21 incomplete or unparseable outputs for Qwen3.5-27B; Gemma-4-12B has no such additional failures.

\par
\WFclear

\begin{findingbox}
\textbf{Finding 4.}
Smaller open-weight models can generate and repair valid policies when the
relevant repository evidence fits within their effective context window, but
their validity is less consistent across languages.
\end{findingbox}

\shortsectionBf{RQ2- Oracle Quality and Coverage:}\label{sec:rq2}
RQ2 evaluates the generated oracles in terms of executability, decision distribution, coverage, and correctness. \emph{Executability} is the proportion of test cases that run successfully in the corresponding policy engine. \emph{Decision distribution} reports the proportions of expected \textsc{Allow} and \textsc{Deny} decisions. \emph{Test-case correctness} is assessed through manual validation.

We measure two types of coverage: \emph{structural coverage} and \emph{condition/path coverage}. Structural coverage measures how much of a policy is exercised by the test cases, with the coverage unit depending on the policy language. For Cedar, the coverage unit is a top-level \texttt{permit}/\texttt{forbid} rule; for OpenFGA, it is a declared \texttt{(type, relation)} pair; and for Rego, it is an executable source line. We use line-level coverage for Rego because authorization behavior can be distributed across multiple rules and helper definitions, making individual rules difficult to treat as independent policy units. A Cedar rule or OpenFGA relation is considered covered if at least one test evaluates it. Rego line coverage is obtained using the Open Policy Agent (OPA)'s built-in coverage analysis.

For condition/path coverage, we measure condition coverage for Cedar and Rego and path coverage for OpenFGA. For Cedar, we consider only rules with a \texttt{when} or \texttt{unless} clause (e.g., \texttt{when \{ principal == resource.owner \}}), and a condition is covered when at least one test evaluates it. For Rego, condition coverage measures the conditions executed by the test cases. For OpenFGA, we define path coverage per relation over its two decision outcomes, \textsc{Allow} and \textsc{Deny}: a relation is
fully covered when test cases observe both outcomes.

\paragraph{Manual validation.} Because test oracle $T_L$ is LLM-generated, its expected decisions are not
assumed to match repository authorization behavior without human validation. We determine the number
of test cases to review using Cochran's
formula~\cite{cochran1977cochran_1977_sampling_techniques_third_} at 95\%
confidence ($z=1.96$), a margin of error $e=0.05$, and the conservative
choice $p=0.5$, giving $n_0=384.16$. We therefore randomly sample 390 test
cases from the generated oracles, with 130 cases from each language. The
margin of error applies to the pooled sample. Each case takes approximately
30--60 seconds to review, for a total manual effort of approximately
3.25--6.5 hours.

\begin{wraptable}{r}{0.52\textwidth}
\vspace{-1.2em}
\centering
\scriptsize
\setlength{\tabcolsep}{3pt}
\begin{tabular}{@{}llrrrr@{}}
\toprule
& & & & \multicolumn{2}{c}{\textbf{Coverage}} \\
\cmidrule(l){5-6}
\textbf{Oracle} & \textbf{Policy} & \textbf{Exec.} & \textbf{Allow} &
\textbf{Struct.} & \textbf{Cond.} \\
\midrule
\multicolumn{6}{@{}l}{\textit{Cedar}} \\
\multirow{2}{*}{GPT-5.4} & Claude S5   & 87.6 & 53.4 & 62.3 & 72.7 \\
                         & DeepSeek V4 & 86.4 & 55.8 & 56.3 & 68.8 \\
\multirow{2}{*}{GLM-5.2} & Claude S5   & 87.9 & 52.9 & 70.2 & 72.5 \\
                         & DeepSeek V4 & 87.7 & 55.1 & 62.7 & 78.0 \\
\midrule
\multicolumn{6}{@{}l}{\textit{OpenFGA}} \\
\multirow{2}{*}{GPT-5.4} & Claude S5   & 97.6 & 48.7 & 67.6 & 65.1 \\
                         & DeepSeek V4 & 99.2 & 49.0 & 78.1 & 77.3 \\
\multirow{2}{*}{GLM-5.2} & Claude S5   & 97.9 & 49.2 & 51.3 & 50.4 \\
                         & DeepSeek V4 & 93.7 & 46.3 & 52.7 & 48.7 \\
\midrule
\multicolumn{6}{@{}l}{\textit{Rego}} \\
\multirow{2}{*}{GPT-5.4} & Claude S5   & 92.0 & 55.2 & 68.1 & 65.8 \\
                         & DeepSeek V4 & 87.8 & 56.3 & 65.8 & 59.8 \\
\multirow{2}{*}{GLM-5.2} & Claude S5   & 92.7 & 52.0 & 62.5 & 55.3 \\
                         & DeepSeek V4 & 86.7 & 41.0 & 60.0 & 53.0 \\
\bottomrule
\end{tabular}
\caption{Test-case executability, share of Allow decisions, and policy
coverage (all in \%) per oracle and policy model pair. Deny is the
complement of Allow. Structural coverage is rule coverage for Cedar,
relation coverage for OpenFGA, and line coverage for Rego; condition
coverage is reported for Cedar and Rego, and path coverage for OpenFGA.}
\label{tab:combined-results}
\vspace{-3em}
\end{wraptable}

Two annotators independently review each test case together with the
supporting evidence in $\mathcal{C}$ and the generated policy $P_L$,
assessing \textit{evidence support} (whether test cases and the expected decision are supported by
the evidence), \textit{policy--test alignment} (whether the test exercises
the intended authorization behavior), \textit{completeness} (whether all
information needed to execute the scenario is present), and
\textit{test-case quality} (e.g., malformed requests, ambiguous evidence, or
unsupported values). A test case is considered correct only if it satisfies all four criteria. Inter-annotator agreement is substantial to almost
perfect, with Cohen's $\kappa$ of 0.762 for Cedar, 0.867 for OpenFGA, and
0.645 for Rego; disagreements are resolved through discussion until
consensus is reached. Of the reviewed cases, 93.1\% for Cedar, 96.9\% for OpenFGA, and 91.5\% for
Rego are correct. Oracle generation is thus not error-free, but
the majority of test cases are supported by repository evidence and
correct.


\textit{Test-case complexity.} Complexity is measured as the number of
co-occurring conditions required to determine an authorization decision.
Cedar and Rego have similar distributions at the upper end, with 36.6\%
and 35.8\% of cases requiring six or more conditions. Below that
threshold, Cedar cases are spread roughly evenly across zero to five
conditions (about 9--16\% each), whereas Rego cases concentrate at two to
four conditions, with fewer than 10\% requiring zero or one. In contrast,
81.9\% of OpenFGA cases require zero or one condition, about 12\% require
two, and fewer than 3\% require more than three. Many OpenFGA decisions
can therefore be determined by individual relationships or shallow
relationship graphs, whereas Cedar and Rego decisions more often depend on
multiple conditions, which is consistent with the agreement differences in
RQ1.



\paragraph{Executability and coverage.}

Table~\ref{tab:combined-results} reports executability, decision
distribution, and coverage for each oracle--policy model combination.
Executability ranges from 86.4\% to 99.2\%. Both decisions are also well
represented: \textsc{Allow} cases account for 41.0--56.3\% of executable
test cases, so neither decision dominates the generated test cases.
Coverage varies across languages and oracle models. For Cedar, rule coverage
ranges from 56.3\% to 70.2\%, while condition coverage ranges from 68.8\%
to 78.0\%. For OpenFGA, relation coverage ranges from 51.3\% to 78.1\%,
and path coverage ranges from 48.7\% to 77.3\%. For Rego, line coverage
ranges from 60.0\% to 68.1\%, while condition coverage ranges from 53.0\%
to 65.8\%. These differences indicate that the two oracle models exercise
different parts of the policies.

\begin{wraptable}{r}{0.42\columnwidth}
\vspace{-1em}
\centering
\scriptsize
\setlength{\tabcolsep}{4pt}
\begin{tabular}{llcc}
\toprule
\textbf{Lang.} & \textbf{Policy} & \textbf{Struct.} & \textbf{Cond./Path} \\
\midrule
\multirow{2}{*}{Cedar}
& Claude S5   & 75.3\% & 76.7\% \\
& DeepSeek V4 & 70.6\% & 85.0\% \\
\midrule
\multirow{2}{*}{OpenFGA}
& Claude S5   & 75.0\% & 73.2\% \\
& DeepSeek V4 & 82.7\% & 81.4\% \\
\midrule
\multirow{2}{*}{Rego}
& Claude S5   & 82.1\% & 72.7\% \\
& DeepSeek V4 & 76.5\% & 65.9\% \\
\bottomrule
\end{tabular}
\caption{Policy coverage when test cases from both oracle models are combined.}
\label{tab:combined-coverage}
\vspace{-3em}
\end{wraptable}

The two oracle models provide complementary coverage. Combining their test
cases increases both structural and condition/path coverage for every
policy model. For Cedar, combined rule coverage reaches 75.3\% for Claude
Sonnet~5 and 70.6\% for DeepSeek-V4-Pro, while condition coverage reaches
76.7\% and 85.0\%, respectively. For OpenFGA, combined relation coverage
reaches 75.0\% and 82.7\%, and combined path coverage reaches 73.2\% and
81.4\%. For Rego, combined line coverage reaches 82.1\% and 76.5\%, and
combined condition coverage reaches 72.7\% and 65.9\%. Thus, although the
two oracle models differ in which parts of a policy they exercise, their
test cases collectively provide broader coverage than either oracle alone.

\par
\WFclear

\begin{findingbox}
\textbf{RQ2 Answer.}
The oracle is reliable and balanced: 91.5--96.9\% of manually reviewed test
cases meet all review criteria, 86.4--99.2\% execute successfully, and
both decisions are well represented. GPT-5.4 and GLM-5.2 cover different
parts of the authorization behavior, and combining their test cases yields
higher coverage than either model alone.
\end{findingbox}



\begin{wraptable}{r}{0.38\columnwidth}
\centering
\scriptsize
\vspace{-2em}
\begin{tabular}{lrrr}
\toprule
\textbf{Language} & \textbf{Total} & \textbf{Valid} & \textbf{Invalid (\%)} \\
\midrule
Cedar   & 3{,}890  & 1{,}723 & 2{,}167 (55.7\%) \\
Rego    & 15{,}131 & 7{,}164 & 7{,}967 (52.7\%) \\
OpenFGA & 360      & 214     & 146 (40.6\%) \\
\bottomrule
\end{tabular}
\caption{Repository policies that pass or fail each language's native validator.}
\label{tab:repository-policy-validation}
\vspace{-3em}
\end{wraptable}

\shortsectionBf{RQ3- Generated vs repository Policies:}\label{sec:rq3}
RQ3 compares the synthesized policies with the repository policies $\mathcal{P}$
both quantitatively and through case studies. We first check whether
$\mathcal{P}$ passes its language's validation tool to assess its suitability
as a baseline. For Cedar, we then run the synthesized and valid repository
policies on the same oracle test cases and measure decision agreement. Finally,
we examine six repositories, two per language, in detail. We compare policies by their behavior, since structurally different policies
can encode the same authorization behavior. The case studies examine
whether a synthesized policy corrects defects, generalizes rules, omits
behavior due to missing evidence, or differs because of repository-specific
execution dependencies.

\paragraph{Validation of Repository Policies}


We assess repository policy validity using the same procedure as for generated
policies, with language-specific validation tools.
Table~\ref{tab:repository-policy-validation}
shows that 55.7\% of Cedar policy statements, 52.7\% of Rego policies, and
40.6\% of OpenFGA models are invalid. Therefore, repository
policies are not reliable ground truth.

\paragraph{Decision Agreement with Repository Policies}
We compare generated and repository policies using the same oracle requests.
We perform this comparison only for Cedar because its schema defines entity
types, attributes, and actions, allowing each request to be evaluated directly
against both policies. For OpenFGA and Rego, generated policies may use
different entity, relation, or input names from the repository policies,
making direct request-level comparison difficult. We first compare complete repository policies. We identify 20 Cedar
repositories in which all repository policies $\mathcal{P}$ are valid, and every oracle request
produces a decision from the repository policy. This allows us to evaluate
the complete authorization behavior encoded by each repository's policies $\mathcal{P}$ as a
whole. We then perform a
broader comparison using individually valid statements across all 53 Cedar
repositories. This includes 1,723 valid statements, with 36 repositories
providing at least one valid repository policy for comparison. The complete-policy comparison achieves 86.5\% agreement between generated and repository policy decisions for Claude Sonnet 5 and 83.9\% for DeepSeek-V4-Pro, based on 1,450 requests. The broader
comparison achieves 76.2\% and 76.0\% agreement, respectively, based on
3,120 requests. We use the complete-policy comparison as the primary measure because it
evaluates each repository policy as a whole. The broader comparison removes
invalid statements before comparison, which can change the policy's overall
authorization behavior because multiple statements may work together to
determine authorization decisions.

\paragraph{Case studies.}

\begin{wraptable}{r}{0.52\columnwidth}
\centering
\scriptsize
\vspace{-2em}
\begin{tabular}{@{}llrrl@{}}
\toprule
 & \textbf{Domain} & \textbf{Stars} & \textbf{Policy Units} &
 \textbf{Policy Validity} \\
\midrule
C1 & Document management   & 9   & 11 & Invalid \\
C2 & Photo-sharing web app & 17  & 7  & Valid \\
\midrule
O1 & Monitoring platform   & 75K & 4  & Invalid \\
O2 & Authorization testbed & 2   & 4  & Invalid \\
\midrule
R1 & Access management layer & 34 & 1 & Valid \\
R2 & GDPR compliance        & 6  & 1 & Valid \\
\bottomrule
\end{tabular}
\caption{Characteristics of the six case-study repositories (C: Cedar,
O: OpenFGA, R: Rego). O1, R1, and R2 are \emph{Production grade} meaning that these are deployed in production.}
\vspace{-2em}
\label{tab:case-study-repos}
\end{wraptable}

Decision agreement does not show \emph{why} generated and repository policies
differ. We therefore conduct case studies on two repositories per language:
Cedar (C1, C2), OpenFGA (O1, O2), and Rego (R1, R2). The cases cover invalid
repository policies, hardcoded rules, and fragmented authorization models.
For each repository, we compare the generated and repository policies in
terms of validity, authorization behavior, and structure, and report decision
agreement.

\begin{wrapfigure}{r}{0.5\columnwidth}
    \centering
    \vspace{-6pt}
    \includegraphics[width=\linewidth]{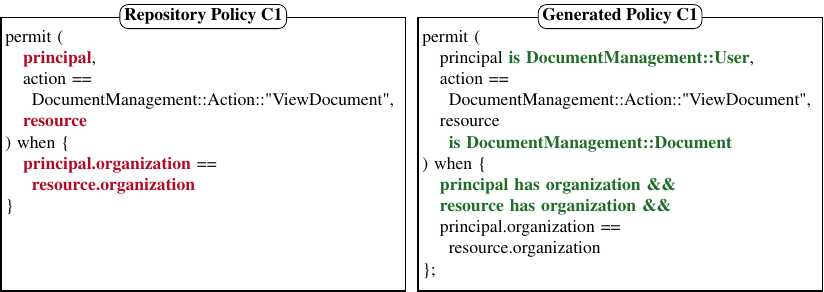}
    \caption{Case C1: the invalid repository policy (left) leaves its entities unscoped and accesses attributes unguarded (red). The generated policy (right) adds type scoping and has checks (green).}
    \label{fig:cases1}
    \vspace{-8pt}
\end{wrapfigure}

\textit{Correcting and generalizing repository policies.}
In C1, all the repository policies are invalid. On the other hand, both generation models produce valid policies that preserve the repository authorization
intent (Figure \ref{fig:cases1}); however, in a more compact form: where the repository states each case as a
separate statement, the generated policies combine alternative conditions
into a single rule using disjunction (\texttt{||}), and therefore contain fewer statements than
the repository's 11: 10 for Claude Sonnet~5 and 6 for DeepSeek-V4-Pro.
For one rule, however, the repository evidence lacks the information needed to
recover the intended logic, and neither model produces it. The generated
policies achieve full rule and condition coverage, with agreement of 26/26
against GPT-5.4 and 24/26 against GLM-5.2 for both models. 

\begin{wrapfigure}{r}{0.50\columnwidth}
    \centering
    \vspace{-6pt}
    \includegraphics[width=\linewidth]{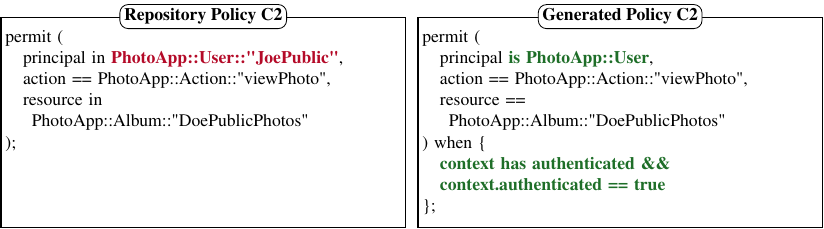}
    \caption{Case C2: the repository policy grants access to one hardcoded user. The generated policy generalizes this to any authenticated user.}
    \label{fig:cases2}
    \vspace{-8pt}
\end{wrapfigure}

In C2, the repository policies are valid but rely on hardcoded principals and
narrow conditions: several inactive policies grant the same
\texttt{viewPhoto} action to individually named users, encoding separate
exceptions rather than reusable logic. Because these principal names are not
available during generation, both models generalize the evidence into
role-, attribute-, and context-based policies. The generated policies achieve
full agreement with both oracles (11/11 with GLM-5.2 and 5/5 with GPT-5.4)
and full rule and condition coverage. Figure~\ref{fig:cases2} illustrates
this generalization: the repository policy grants \texttt{viewPhoto} on
\texttt{DoePublicPhotos} to the single user \texttt{JoePublic}, whereas the
generated policy generalizes the principal to any \texttt{PhotoApp::User}
while allowing the exact album \texttt{DoePublicPhotos}. It also requires an authenticated session,
guarded by a \texttt{has} check, thereby extending access to the intended
population without expanding the set of protected resources.



\textit{Equivalent intent, different structure.}
R1 is a valid production Rego policy for identity and access management.
Both models recover its core authorization logic but use different
structures. Claude S5 introduces an explicit default-deny rule, more detailed
administrator checks, and action categories, while DeepSeek-V4-Pro produces a
more compact policy. Agreement varies across the two oracles: Claude S5
achieves 8/8 with GLM-5.2 and 38/51 with GPT-5.4, while DeepSeek-V4-Pro
achieves 22/23 and 19/26, respectively. R2 implements GDPR-related role-based access control with a default-deny
policy. Both models recover the core structure, using role-based authorization
for operations with a default-deny rule, but express it differently from the
repository policy. Claude S5 achieves 12/15 agreement with GPT-5.4 and 8/8
with GLM-5.2, while DeepSeek-V4-Pro achieves 6/10 and 10/13, respectively.


\textit{Configuration and fragmentation.}
In O1, the repository policy consists of four \texttt{.fga} files that fail
validation; adding a minimal \texttt{fga.mod} file restores validity without
modifying them, so the failure stems from configuration rather than
authorization logic. The generated policies recover the relationships of
one module, whose structure is well represented in the retrieved evidence,
but not of the remaining modules, whose evidence falls outside the retrieved
context. Agreement is high (Claude Sonnet~5: 34/34 with GPT-5.4 and 23/24
with GLM-5.2; DeepSeek-V4-Pro: 32/32 and 26/26), but relation coverage is
9/14 and 11/14, respectively. O2 also
contains four OpenFGA files. The \texttt{fga.mod} fix from O1 is insufficient
here: one file is repaired with minor syntax fixes, while the other three
depend on type information missing from the repository and are left
unrepaired rather than completed with unsupported logic. These files still
describe the application's entities and relationships, and the generated
policy covers all these information from the evidence. Agreement is
complete for all model--oracle pairs (Claude Sonnet~5: 56/56 and 29/29;
DeepSeek-V4-Pro: 33/33 and 31/31), as is structural coverage (26/26
relations and 52/52 paths, and 28/28 and 56/56, respectively). Generated
policies can therefore recover a more complete authorization model than any
single fragmented repository file when the intent is distributed across the
repository.

\begin{findingbox}
\textbf{RQ3 Answer.}
Generated policies match the decisions of completely valid repository Cedar
policies on 83.9--86.5\% of oracle requests. The case studies further show that
\Sys can generate authorization behavior when repository policies are invalid,
fragmented, or structured differently, and can correct defects and generalize
hardcoded rules when sufficient evidence is available.
\end{findingbox}

\begin{wraptable}{r}{0.38\columnwidth}
\centering
\scriptsize
\setlength{\tabcolsep}{3pt}
\begin{tabular}{llrrr}
\toprule
\textbf{Stage} & \textbf{Model} & \textbf{Tokens} & \textbf{Cost} & \textbf{Time} \\
\midrule
\multirow{2}{*}{Policy}
& Claude S5   & 232.5K & \$0.58 & 1.02 min \\
& DeepSeek V4 & 158.5K & \$0.08 & 1.05 min \\
\midrule
\multirow{2}{*}{Oracle}
& GPT-5.4 & 308.5K & \$0.99 & 3.75 min \\
& GLM-5.2 & 387.7K & \$0.47 & 4.43 min \\
\bottomrule
\end{tabular}
\caption{Average tokens, API cost, and time per repository for policy and oracle generation.}
\label{tab:computational-cost}
\vspace{-8pt}
\end{wraptable}
\shortsectionBf{Computational Cost:}
\label{sec:computational-cost}
Table~\ref{tab:computational-cost} reports average token consumption, API
cost, and execution time per repository over the 228 repositories.
API costs use the providers' published
prices from August~2026. A complete policy generation and oracle assessment takes about five minutes
and costs \$0.55--\$1.57, depending on the model pair. The smaller open-weight
models run on a university HPC cluster without API charges; Qwen3.5-27B and Gemma-4-12B take around 12 and 3 minutes per repository, respectively.

%% file: Content/background.tex
\shortsectionBf{Policy Synthesis from Requirements.}
A substantial body of work derives access-control policies from artifacts that
explicitly state authorization intent. Prior studies ~\cite{xiao2012automated, slankas2013access,slankas2014relation,narouei2018automatic} extract
authorization statements from natural-language software documents and
translate them into formal policies,
while related work starts from user stories~\cite{heaps2021access}. Recent
LLM-based approaches perform policy identification, attribute extraction, and
generation from natural-language specifications, access requirements, and
unstructured text~\cite{jayasundara2024sok,shan2024automatic,lawal2024translating,vatsa2025synthesizing,mai2025llm,sonune2025lmn,sonune2026laps,paratore2025plain,jayasundara2026agentvlm,vatsa2026autocedar,gupta2026prose2policy,cheng2025say},
including retrieval-augmented generation~\cite{jayasundara2024ragent}.
These approaches assume that authorization intent is explicitly specified in
the input, whereas \Sys does not depend on such specification.

\shortsectionBf{Policy Mining and Reverse Engineering.}
Policy mining infers RBAC, ABAC, or ReBAC policies from structured
authorization data, access logs, or relationship graphs~\cite{xu2014mining,cotrini2018mining,bui2019greedy,bui2020decision}. Recent LLM-based approaches mine
policies from attribute data or access logs~\cite{bui2026mining,babasaheb2025towards}.
Reverse-engineering systems instead recover authorization behavior from
applications, including Android middleware, running web applications, and
microservice interactions~\cite{gorski2019acminer,le2022automated,li2021automatic}.
Program specification systems such as SpecGen infer formal specifications from
code~\cite{ma2025specgen}. Unlike these approaches, which primarily rely
on a single artifact and often evaluate against a reference policy,
\Sys combines code, comments, and documentation and evaluates generated
policies independently using oracles.

\shortsectionBf{Policy Testing.}
Access-control testing has used policy-specific coverage criteria and
model-based test generation~\cite{martin2006defining,xu2012model,xu2018automated,xu2020automated},
typically deriving tests from the policy or an explicit access-control model.
LLMs have also been applied to test generation~\cite{tang2024chatgpt,dakhel2024effective},
or validate generated
policies against requirements-derived references~\cite{vatsa2026autocedar}.
These approaches assume a policy or specification that can serve as
a trusted reference, whereas \Sys derives its test oracle
from repository evidence.

%% file: Content/discussion.tex
\section{Limitations}
\shortsectionBf{Dataset and generalizability.}
Our dataset contains 228 public repositories and may not represent private
codebases or other application domains. \Sys also depends on language-specific
artifacts: a schema for Cedar and repository policy files for retrieving
evidence in OpenFGA and Rego. The evaluated models may have seen some of
these public policies during training. 

\shortsectionBf{Evaluation scope.}
Agreement with the generated oracle does not necessarily reflect the
application's true intended policy. The oracle uses a finite set of generated
requests and may miss some behaviors. For OpenFGA and Rego, requests are
created from the generated policy interface, so behavior missing from that
interface cannot be tested. Retrieval may also miss relevant evidence,
especially when authorization logic is implicit in the code. The retrieval
threshold $\tau$ was selected using a six-repository pilot, and manual
validation may introduce evaluator subjectivity.

The oracles are generated by LLMs and may contain incorrect decisions. We
reduce this risk by using different models, requiring supporting evidence
for each decision, and manually checking the generated tests. Because LLM
outputs can vary across runs, the results may also vary.

%% file: Content/conclusion.tex
\Sys synthesizes access-control policies from code, comments, and
documentation, repairs them with native validator feedback, and validates
their behavior against a behavioral oracle derived independently from the
same evidence. On 228 real-world repositories spanning Cedar, OpenFGA,
and Rego, \Sys produces 99--100\% valid policies, and 91.5--96.9\% of
manually reviewed oracle test cases are correct. Decision agreement is
highest for OpenFGA and lower for Cedar and Rego, whose decisions combine
more conditions. Our case studies show that, when repository evidence is
sufficient, \Sys can correct defects, generalize hardcoded rules, and
recover authorization behavior that invalid or incomplete repository
policies fail to capture.